# What happens when a journal converts to Open Access?
# A bibliometric analysis

Fakhri Momeni[1*], Philipp Mayr[1,2*], Nicholas Fraser [3*] and Isabella Peters[4]

[1] firstname.lastname@gesis.org
GESIS – Leibniz Institute for the Social Sciences, Unter Sachsenhausen 6-8, 50667 Cologne (Germany)
[2] mayr@informatik.uni-goettingen.de
Institute of Computer Science, University of Göttingen, (Germany)
[3] N.Fraser@zbw.eu, [4] I.Peters@zbw.eu
ZBW – Leibniz Information Centre for Economics, Düsternbrooker Weg 120, 24105 Kiel (Germany)

* Correspondence: fakhri.momeni@gesis.org, N.Fraser@zbw.eu and philipp.mayr@gesis.org

## Abstract

In recent years, increased stakeholder pressure to transition research to Open Access has led to many journals converting, or 'flipping', from a closed access (CA) to an open access (OA) publishing model. Changing the publishing model can influence the decision of authors to submit their papers to a journal, and increased article accessibility may influence citation behaviour. In this paper we aimed to understand how flipping a journal to an OA model influences the journal's future publication volumes and citation impact. We analysed two independent sets of journals that had flipped to an OA model, one from the Directory of Open Access Journals (DOAJ) and one from the Open Access Directory (OAD), and compared their development with two respective control groups of similar journals. For bibliometric analyses, journals were matched to the Scopus database. We assessed changes in the number of articles published over time, as well as two citation metrics at the journal and article level: the normalised impact factor (IF) and the average relative citations (ARC), respectively. Our results show that overall, journals that flipped to an OA model increased their publication output compared to journals that remained closed. Mean normalised IF and ARC also generally increased following the flip to an OA model, at a greater rate than was observed in the control groups. However, the changes appear to vary largely by scientific discipline. Overall, these results indicate that flipping to an OA publishing model can bring positive changes to a journal.

**Keywords**: Open Access Publishing, Scholarly Communication, Citation Analysis, Scholarly Journals, Journal Publishing Models.

## Introduction

For hundreds of years, the closed-access (CA) model has been the traditional publishing model, where journal articles are published behind a "paywall" that can be removed through the payment of subscription fees to the publisher, most commonly by academic libraries or research funders. Over the past three decades, the growth of the Internet and resulting opportunities for low-cost distribution of digital content have led to a revolution in scholarly publishing (Laakso et al., 2011; Björk, 2017). In the midst of these changes, a new business model for publishers of scholarly journals has emerged besides the traditional model: an open-access (OA) model, where journal articles are made freely-available to all readers, and the publication costs are borne by third-parties, such as authors, institutions, societies or funders. These publishing models may not be mutually exclusive (e.g. a CA journal may still allow certain articles to be published under OA licenses, usually referred to as "Hybrid" publishing), and may not remain static over time; a journal may convert from a CA model to an OA model or vice versa, processes commonly termed as "flipping" or "reverse flipping", respectively (Solomon, Laakso, & Björk, 2016; Matthias et al., 2019). Recent quantitative studies found e.g. that more than 50% of the newer articles indexed by Web of Science are freely available in "some form" of OA via Google Scholar (Martin-Martin et al., 2018). As a result, the speed of adoption of OA is increasing constantly. Hobert et al. (2020) can show this trend (OA uptake) in a large-scale study for German universities and non-university research institutions in the period 2010-2018. They found out that 45% of all considered articles in the observed period 2010-2018 were openly available at the time of analysis in one form of OA (Green OA, Gold OA and other OA variants). Hobert et al. showed for Germany that subject-specific repositories are still the most prevalent OA type, but fully OA journals are steadily increasing in the analysed period.

Journal flipping in itself is not a new concept: Peter Suber previously noted that "*Subscription journals have been converting or "flipping" to open access (OA) for about as long as OA has been an option*" (in Solomon, Laakso, & Björk, 2016). However, the topic has received more attention in recent years due to increased funder pressure to accelerate the transition to OA, for example through Plan S in Europe (https://www.coalition-s.org/), driven in part by the increasingly unsustainable economic costs of the subscription model (Schimmer et al., 2015; Tennant et al., 2016). For publishers who intend to transition from a CA to an OA business model there is an urgent need to understand the journal flipping process and its consequences. A clear concern of these publishers is to find an alternative stable stream of income to subscription fees. OA journal revenue streams are commonly associated with Article Processing Charges (APCs), whereby authors, institutions or funders pay fees to a publisher on a per-published-article basis. Björk (2012) demonstrated that this author-pays model in hybrid journals is unpopular with authors, whilst Peterson et al. (2013) argued that the cost of APC in this model is often too much for many authors, and publishers try to solve this problem in different ways such as fee waiver policies, subsidizing academic publishing directly without profiteering intermediaries, etc. However, according to the public journal dataset from the Directory of Open Access Journals (DOAJ)[1], only 4,021 of 14,741 (27.2%) of journals charge APCs (data accessed on 10th June 2020); the remainder may be supported, for example, by individual societies or library presses. Even so, according to Solomon, Laakso, & Björk (2016), transitioning a journal to an OA model for those societies with low numbers of publications can be expensive.

Predicting how a change in the business model will affect the long-term viability of a flipped journal is of immense importance to those responsible for journal management, thus in-depth, longitudinal bibliometric studies can help to inform decision making of publishers, and their assessment of chances and risks of flipping their journals (see e.g. Perianes-Rodríguez & Olmeda-Gómez, 2019). Such bibliometric studies may focus on multiple aspects of publishing behaviour, such as changes in publishing volume, which is itself a function of submission volumes and editorial selection processes, as well as changes in article impact, which may be a proxy for the future "attractiveness" of a journal to researchers. This study addresses both of these aspects, building on work presented in Momeni et al. (2019) but substantially expanding its scope, in terms of the data sources of flipped journals and the bibliometric data analysed (from Web of Science to Scopus). Moreover, in this study we included a comparison of flipped journals and journals from the same disciplines that still publish under the CA model (as suggested recently by Bautista-Puig et al., 2020). We aim to answer the following research questions:

(1) Do journals flipping from a CA to an OA model experience a positive/negative change in the volume of articles published?
(2) Do journals flipping from a CA to an OA model experience a positive/negative change in their Impact Factor?

---

[1] https://doaj.org/, public metadata dump available at https://doaj.org/public-data-dump

(3) Do articles in journals flipping from a CA to an OA model experience a positive/negative change in their individual citation impact?

An important point to note, is that this study focuses only on journals that have flipped from a CA to an OA model, whilst retaining the same journal name. Over the past years a number of journal "declarations of independence"[2] have resulted in the resignation of editors from a CA journal to form a new OA journal at an alternative publisher (e.g. the editorial board of *Journal of Informetrics*, published by Elsevier, transitioned to a new journal called *Quantitative Science Studies*, published by MIT Press; Waltman et al., 2020). Although these transitions are closely related to the concept of "flipping", they not only concern the journal name, but also involve a change in journal venue and potential attractiveness, which may make the direct effects of transitioning from CA to OA difficult to distinguish. In our study we considered just journals which kept the same journal name after flipping.

**Related work**

*Studies on journal flipping from a bibliometric perspective*

Relatively few studies have been carried out that systematically research the effects flipping has on a journal's publication output and impact. One of the earliest studies from Solomon et al. (2013) documented the growth of OA journals, their articles and normalized citation rates between 1999 and 2010, whilst also controlling for whether the journal had been launched as an OA journal, or flipped to an OA journal at a later point. The authors combined data from Scopus and DOAJ, and manually reviewed the public websites of all journals included in their matched dataset (N = 2,012), finding that of these journals, 1,064 were flipped from a CA to an OA model, whilst 931 journals were "born-OA" (17 were undetermined). According to the data from Solomon et al. (2013), the number of flipped OA journals peaked in 2005, and since then decreased year-on-year; in 2012 less than 20 journals were discovered that had flipped from a CA to an OA model. In terms of citations, the authors compared longitudinal trends in Source Normalized Impact per Paper (SNIP), a citation metric that accounts for field-specific differences in citation. They found that overall citation rates for flipped OA journals were approximately 50% lower than those from CA journals, but this relationship did not change greatly over time. Conversely, born-OA journals experienced a strong growth in SNIP between the years 2003-2005, eventually reaching a plateau with citation rates almost at the same levels of CA journals.

In another study, Busch (2014a, 2014b) investigated the response of the Impact Factor (IF) of six journals which were transferred from CA models at other publishers to the OA model of BioMed Central between the years 2006 and 2011. IFs were compared to the median IF of journals from the same Web of Science subject category. Four of the six journals experienced a sizeable increase in IF following the flip to OA - for example the *Journal of Cardiovascular Magnetic Resonance* increased its IF from 1.87 in the year prior to flipping, to 4.33 in the year after flipping, a ~130% increase. For the remaining two of the journals, IFs remained relatively static or even fell following the flip, although the author notes that the goal of these journals for the years in question was to increase their publishing volume, which may have led to less selective editorial decisions; both journals in fact accepted around 50% more submissions in the post-flip years than pre-flip. These results must, however, be interpreted carefully; not only did the journals flip from a CA to an OA model, but they also transferred to a new publisher (although keeping their old name) which may have had an important effect on the journal's visibility.

As well as converting from a CA model to an OA model, some journals may also convert in the opposite direction, from an OA to a CA model, a process that has been termed "reverse flipping" (Matthias et al., 2019). The study of Matthias et al. (2019) investigated the publication and citation behaviour of 152 journals that were identified as having reverse-flipped from 2005 onwards. Interestingly, 62% of those journals had initially been CA journals and flipped to an OA model, before flipping back to a CA model. The authors also found that publication volumes and citation metrics changed little in the two years before or after the reverse flip, although some individual journals encountered large variability. Reasons for reverse flipping may in part be due to a lack of success with the OA model, for reasons such as financial sustainability or low article volumes, although 69% of reverse flips were related to a change in publisher and thus the journal may have simply adopted the prevalent publication model of the new publisher.

A more recent study by Bautista-Puig et al. (2020) follows a similar methodology to this study. The authors used data combined from DOAJ (N = 119 journals) and the Open Access Directory (OAD)[3] (N = 100 journals), who host a community-maintained list of journals that have flipped from a CA (referred to by OAD as "TA", for Toll Access) to an OA publishing model[4]. The authors compared post-flip and pre-flip bibliometric indicators including publishing volumes and normalised citation rates, against two distinct control groups: a standard control group, as well as a "tailor-made" control group accounting for a journal's national orientation. The authors found evidence of an OA citation advantage: DOAJ journals increased their normalised IF by ~50% at 4-years post-flipping, compared to just ~10% in the standard control group, whilst OAD journals increased their normalised IF by ~35% in the same time interval, compared to ~15% in the standard control group. However, the authors found no evidence for an OA publication advantage: for all groups, the journals experienced an increase in publishing volumes in the range of 10-20%. The authors also assessed changes in the affiliation countries of publishing and citing authors after a journal had flipped. They found that overall, the share of authors from high-income countries declined after a journal flipped to an OA model, although a similar effect was also found in the respective control groups.

The present study is an extension of the previous study of Momeni et al. (2019). In the previous study, we used a list of flipped journals available from OAD, as also used by Bautista-Puig et al. (2020). The list of journals was matched to journals contained in the Web of Science (N = 171) to determine the effects on publication volume and two citation metrics, one at the journal level (IF) and one at the article level (average of relative citations; ARC), of flipping a journal to OA. These initial results showed that flipping a journal mostly had positive effects on a journal's IF, but conversely had no strong effect on the citation impact at the level of individual articles. We also observed a small decline in the number of articles that were published by a journal after flipping to an OA model. Whilst these initial findings were interesting, they also came with several limitations: (1) we only considered a small sample size of journals from a single source, (2) we did not consider the relevant journal and article metrics with respect to any form of control group, thus we could not interpret whether these changes deviated from global publishing and citation patterns, and (3) we did not consider how publication and citation behaviour might vary in different scientific communities. We therefore attempt to address these

---

limitations in the current study, by increasing our sample size with the addition of a new list of journals from DOAJ, by generating a control group for comparing to the group of flipped journals, and conducting analysis at the level of scientific disciplines.

*The OA citation advantage*

In our study we also aim to report on changes in citation impact resulting from a journal flipping from a CA to an OA model, both at the journal level and article level. A number of studies have already attempted to study the relationship between OA and citation impact, with most evidence pointing towards an open access citation advantage (OACA) for OA articles over CA articles (Swan, 2010; Piwowar & Vision, 2013; Sotudeh et al., 2015; McKiernan et al., 2016; Lewis, 2018, Ottaviani, 2016). The Scholarly Publishing and Academic Resources Coalition (SPARC) maintained a repository of 70 studies investigating the OACA[5] until 2015; of these, 46 (65.7%) found a citation advantage, whilst only 17 (24%) found no advantage (the remaining 7 records were inconclusive). A subsequent large-scale study of 3.3 million articles published between 2007 and 2009 by Archambault et al. (2016) found that OA papers received ~23% higher citation impact overall than the global average citation rate, although the effect was stronger in Green OA (i.e. OA articles made available through open repositories) forms than Gold OA (OA articles published in fully OA journals). These findings were echoed in a study by Piwowar et al. (2018), who found that OA articles receive on average ~18% more citations than CA articles, but again this advantage was driven primarily by Green OA, whilst Gold OA was found to have slightly lower citation rates than the global average. Whilst many of these studies note a strong correlation between OA and citation rates, it is important to note that these findings do not necessarily imply causation, as citations may be influenced by a number of additional structural and author-specific factors (Tahamtan, Safipour, & Ahamdzadeh, 2016). Other studies based on randomized control trials (Davis, 2011) have also reported conflicting results, indicating that methodologies taking into account multiple factors are necessary to understand the exact mechanism driving higher citation rates of OA articles.

## Data and methods

*Groups of flipped journals*

For this study we compiled groups of flipped journals from two main sources: DOAJ and OAD. DOAJ is a directory of ~14,700 OA journals, maintained by the Infrastructure Services for Open Access (IS4OA). Journals must apply for indexing in DOAJ and meet a set of basic quality control and transparency criteria to be included. DOAJ provides access to metadata of all indexed journals, which includes a field containing the first calendar year that a complete volume of the journal provided OA to the full text of all articles (herein referred to as "flipping year"). Note that journal metadata in DOAJ is provided by the publishers directly and is thus not "verified" by any third party. As Sotudeh and Horri (2007) and Bautista-Puig et al. (2020) have shown this can often lead to inaccurate data, e.g., in terms of flipping date. To build a group of flipped journals, we extracted details of all journals in DOAJ as well as the flipping year. This group of journals was compared to the *Zeitschriftendatenbank* ("Journal database"; ZDB[6]), a database of high-quality journals and other periodicals maintained by the *Staatsbibliothek zu Berlin* ("State Library of Berlin"). An advantage of using the ZDB is that they maintain a "first issued year" field for each contained journal, and thus by comparing this year with the flipping year field from DOAJ, we can discover journals that were previously a CA journal and then changed to an OA model (i.e. we exclude any journals that were initiated as OA journals). For bibliometric analysis, this group of journals was then matched to journals contained in Scopus via matching of journal names (case-insensitive) and ISSNs. We therefore have only considered journals that had the same names and ISSNs before and after the flip. Access to Scopus was provided via the German Competence Centre for Bibliometrics[7], who maintain an in-house, quality-controlled version of the Scopus database. To follow common standards of bibliometric studies, we applied a number of filters to journals matched between the datasets, namely that the journal must have flipped between 2001 and 2013, that there must be more than 5 years distance between the first issued year and the flipping year, and the journals must have published citable articles in every year for the 4 years prior to and following the flipping year. This resulted in a final group of 234 flipped journals from DOAJ.

---

[5] https://bit.ly/SPARC-OACA

[6] https://www.zeitschriftendatenbank.de

The second group of journals was derived from OAD, a wiki where the OA community can create and support simple factual lists about open access to science and scholarship, hosted by the School of Library and Information Science at Simmons College. OAD contains a community-maintained list of journals that have flipped from CA to OA. We manually retrieved the full list of journals as well as their flipping years from the public web page. Annotations on the website described whether the journal had flipped to a full OA or a hybrid OA model - in this study we only retained journals that flipped to a full OA model. The group of journals were matched to journals indexed in the Scopus via matching of journal names. Just journals with citable articles in all four years around the flipped year, and with flipping years between 2001 and 2013 were included in the study. The final OAD-group contains 87 journals.
Our two compiled groups of journals from DOAJ and OAD have 12 journals in common. In the following, we will treat these two groups as independent journal groups[8].

*Control groups*

For comparative analysis we defined a control group of CA journals for each of the two groups of flipped journals. The control journals were designed to be similar to our flipped journals in terms of discipline, number of published articles and IF in the year of journal flipping. We first defined a candidate list of CA journals, which were obtained from data in Unpaywall, a service that finds OA versions of journal articles and also provides open access to metadata relating to journal publishing models. We used the metadata fields "journal_is_oa" and "article_is_oa" to generate a list of CA journals that do not contain any OA articles (i.e. where journal_is_oa = FALSE and article_is_oa = FALSE for all articles within a journal). These journals were matched with journals contained in Scopus on the basis of shared journal titles (case-insensitive), and then for each journal in our groups of flipped journals, the top 20 percent of CA journals were taken from the same discipline, with the smallest difference in number of published articles in the flipping year. Lastly, from this group of journals with similar volume of articles, a single control journal was selected with the smallest difference in calculated IF to the flipped journal in the flipping year. For the journals with multiple disciplines, a control journal was selected from each individual discipline of the flipped journal. With this method, we have generated two separate control groups, one for the list of flipped DOAJ journals and one for the flipped OAD journals. However, these two control groups may entail common journals.

*Bibliometric indicators used in study*

In this study, we investigate the effect of flipping from CA to OA through a descriptive analysis of the timeline of CA to OA conversions, the change in the number of articles published by the flipped and control journals over time, as well as two metrics of citations at the article and journal level: the average relative citations (ARC) and normalised impact factor (IF), respectively. An explanation of the latter two metrics is given in the following paragraphs.

ARC is calculated, by first calculating a relative citation (RC) count for each individual article published within our flipped journal and control journal datasets, normalised to account for different citation patterns across disciplines. For this calculation we only included articles with the "type" property of "Article" or "Review", as contained within Scopus. The RC of a paper is calculated for each year by computing the sum of citations gained by the individual article, divided by the average number of citations of all papers across its discipline(s) published in the same year. We use a citation window of three years. An RC value above 1 means that a paper is cited more frequently than the average citation level for all papers in that discipline, and vice versa. To calculate the citation performance of a group of papers relative to papers in the same discipline and publication year, we simply calculate the arithmetic mean of the RC of all papers in the group, referred to as the average of relative citations (ARC).

| Fields | Disciplines |
|---|---|
| Physical Sciences | Chemical Engineering<br>Chemistry<br>Computer Science<br>Earth and Planetary Sciences<br>Energy<br>Engineering<br>Environmental Science<br>Material Science<br>Mathematics<br>Physics and Astronomy<br>Multidisciplinary |
| Health Sciences | Medicine<br>Nursing<br>Veterinary<br>Dentistry<br>Health Professions<br>Multidisciplinary |
| Social Sciences | Arts and Humanities<br>Business, Management and Accounting<br>Decision Sciences<br>Economics, Econometrics and Finance<br>Psychology<br>Social Sciences<br>Multidisciplinary |
| Life Sciences | Agricultural and Biological Sciences<br>Biochemistry, Genetics and Molecular Biology<br>Immunology and Microbiology<br>Neuroscience<br>Pharmacology, Toxicology and Pharmaceutics<br>Multidisciplinary |

**Table 1. Fields and disciplines as provided by Scopus.**

For each journal in our dataset, we calculated the two years IF following a similar methodology to that commonly associated with the Journal Citation Reports, produced by Clarivate Analytics[9]. Based on this definition, the IF is defined as all citations to the journal in the current year to items published in the previous two years, divided by the total number of citable items (these comprise articles, reviews, and proceedings papers) published in the journal in the previous two years. In order to compare IFs between different disciplines, we conducted an additional normalisation step using the rescaling method introduced by Radicchi, Fortunato and Castellano (2008). So the citation rate for each individual article used in the IF calculation was rescaled by dividing by the arithmetic mean of the citation rate of all articles in its discipline.

To calculate RC and normalized IF across disciplines we used the "All Science Journal Classification" (ASJC) classification system[10] of Scopus which has four fields (called 'subject areas' in Scopus) and 27 disciplines (called 'subject area classifications' in Scopus; see Table 1). In this classification system, journals can have more than one category, therefore we considered the mean citation rate of all articles in all disciplines of which the journal belongs to.

**Results**

In the following we will present the results of our descriptive analysis.

*Analysis of the flipping year*

IFs are calculated based on citations earned by articles published in the two past years, thus we expect to observe the impact of converting to OA at least one year after the flip. Due to the journal review time, e.g. when a journal flips, newly submitted articles will take several months to proceed through the review process. Therefore for the OAD-group (in the case of having the month of flipping) we considered the following year as the flipping point for journals which were flipped in the fourth quarter to ensure that articles reflect the OA model under which they were submitted. Figure 1 shows the distribution of years in which journals in the two datasets flipped. We observe a peak in the number of flipped journals in 2006, as well as a long-term steady increase in the number of journals that have converted to OA across all years. The peak in 2006 for OAD is caused by a large number of journal conversions carried out by two major publishers: *CSIC Consejo Superior de Investigaciones Científicas* and *Hindawi*. The peak for the DOAJ journals in 2013 has different publishers and is not dominated by specific publishers. Figure 2 shows the distribution of flipped journals across fields and disciplines. The majority of journals are categorised into the disciplines 'Medicine', 'Social Sciences', 'Agricultural and Biological Sciences' and 'Arts & Humanities'. However, the two groups differ with regard to disciplines included: the OAD-group seems to include a considerably higher amount of journals from 'Arts & Humanities' and 'Mathematics' than the DOAJ-group.

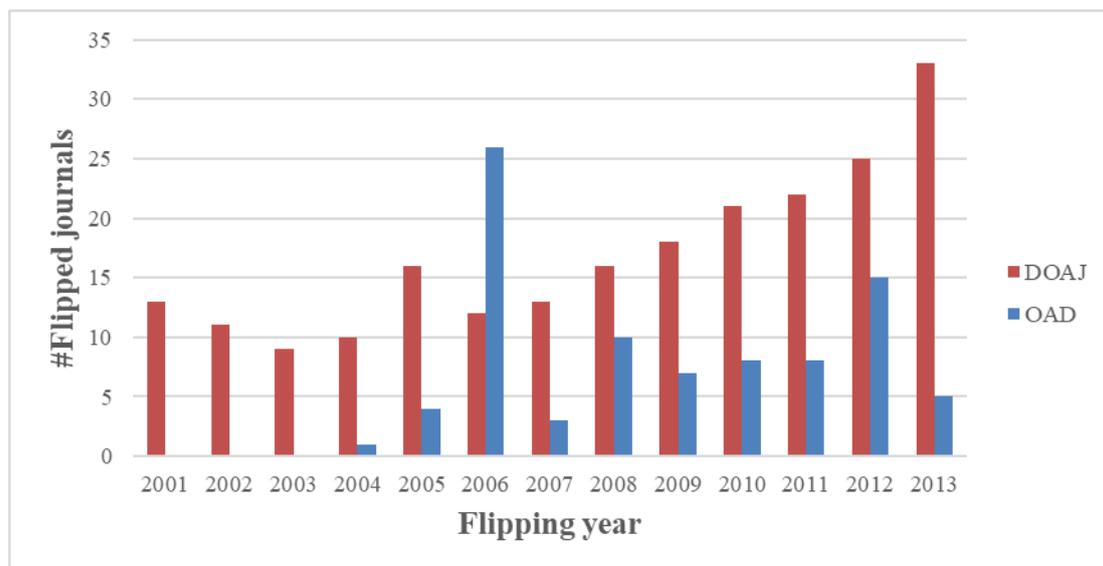

**Figure 1. Distribution of flipped DOAJ and OAD journals by year. In this study we considered only journals with flipping years between 2001 and 2013.**

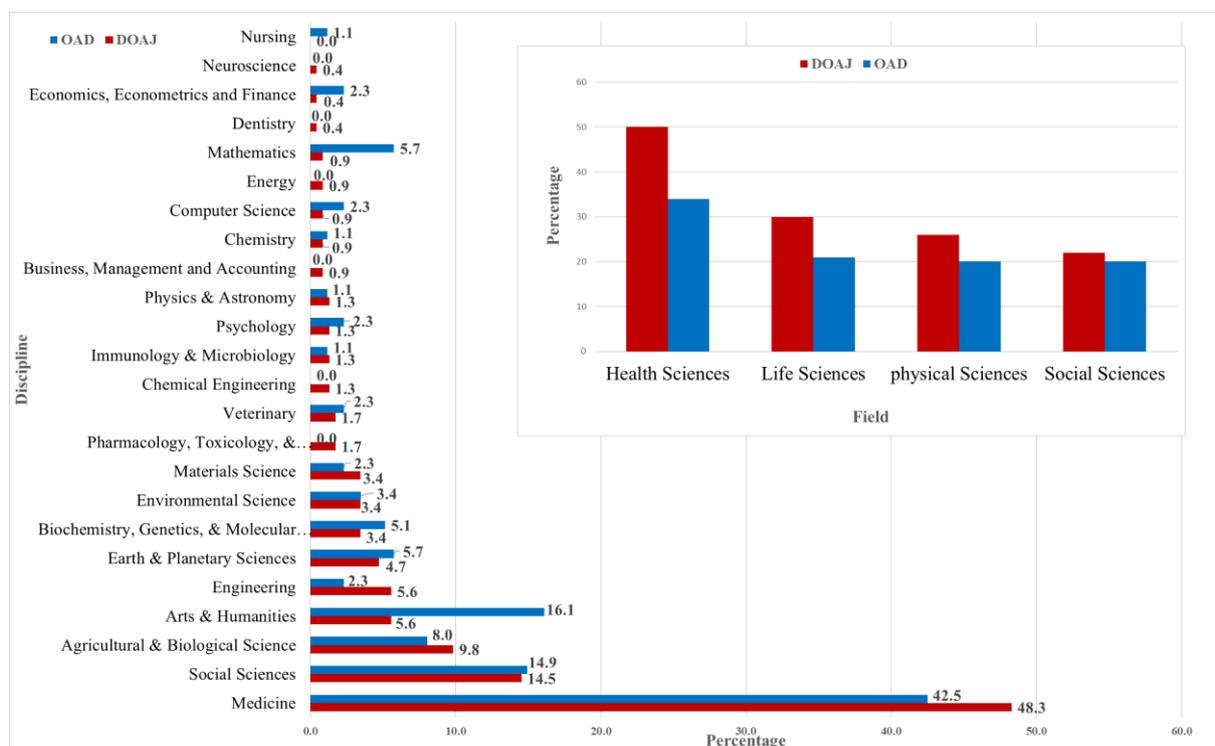

**Figure 2. Proportion of flipped DOAJ and OAD journals per discipline and field. Note that a journal can belong to more than one discipline; thus, percentages do not sum up to 100%. Only disciplines that included at least one journal are shown.**

*Journal publishing volumes*

We assessed the evolution of publishing volumes for journals that flipped from a CA to an OA model, for 4 years prior to and 4 years following the year of the flipping (see Figure 3). For each group of flipped journals (i.e. DOAJ or OAD) we calculated the mean number of articles published per journal per year and compared these numbers to the control group. Independent from the flipping process, we can observe both groups of flipped journals publish fewer articles on average than their respective control.

The number of articles published in the flipping year ranged from 1 in *Journal Hungarian Geographical Bulletin*, to 3,807 in *Journal of Acta Crystallographica Section E*. In general, we observe a small but steady increase in the number of articles published by journals following the flip to an OA model, which continues for the entire 4-year period of our analysis. For DOAJ flipped journals, the mean number of published articles increased from 80 articles in the flipping year, to 98 articles 4 years after flipping, an increase of 22.5%. In contrast, the DOAJ control group only increased from 59 to 67 articles on average, an increase of 14%. For OAD flipped journals the number increased from 112 articles in the flipping year, to 128 articles 4 years after flipping, an increase of 14.3%, whilst the control group decreased from 104 to 103 articles in the same period, a decrease of 1%. Thus, for both groups of journals the mean increase in publishing volumes for flipped journals exceeded the increase in publishing volumes for journals that remained CA.

For OAD flipped journals, the post-flip increase in publishing volume appears to be insensitive to general long-term trends, as in all three years prior to flipping the number of published articles per journal remained relatively static at ~111 articles per year, and only began to increase prominently at 2 years post-flipping. For DOAJ, the interpretation is less clear – in general the number of published articles increased in the 4-year period prior to flipping, but the trend is characterised by a decline in the number of published articles in the year immediately preceding the flip.

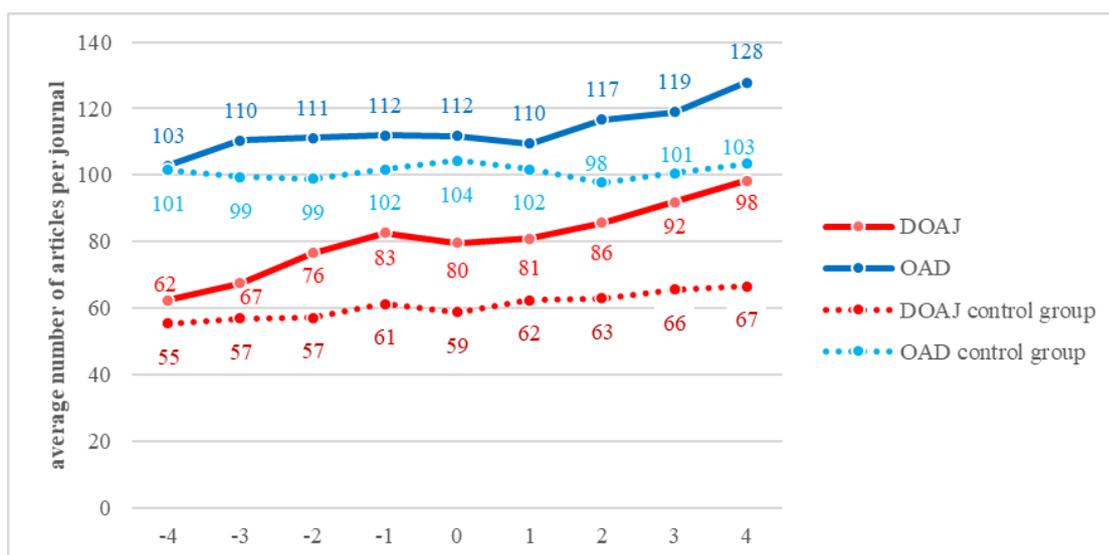

**Figure 3. Yearly average number of articles for flipped journals four years before and after flipping. The x-axis refers to the year with respect to the flipping year: 0 represents the year of flipping, positive values the years following the flip, and negative values the years preceding the flip.**

*Article and journal level citation metrics*

Figure 4 shows the mean normalised IF (red line) and ARC (blue line) for journals and articles, respectively, in our dataset for the four years before and after flipping. The ranges of IF and ARC in the year of flipping are from 0 to ~5.55 and 0 to 60.8 respectively. The left panel shows the values for journals and articles within the OAD flipped journals (solid line) and respective control group (dashed line), and the right panel the same for DOAJ journals. To also more clearly demonstrate changes in ARC and normalized IF at specific time points following flipping, we also calculate growth rates of each metric at two and four years post-flipping, relative to the values in the flipping year (see Figure 5). For OAD flipped journals, we observe no major difference in ARC after flipping, as values remain relatively stable in the range from ~1.8 to ~1.9. In terms of normalized IF, we observe a small increase for OAD journals, from 0.69 in the flipping year to 0.82 at 4 years after flipping (an increase of 19%). However, this increase is relatively small and not significantly greater than the interannual variability that we observe in either the OAD flipped journals or the respective control group.

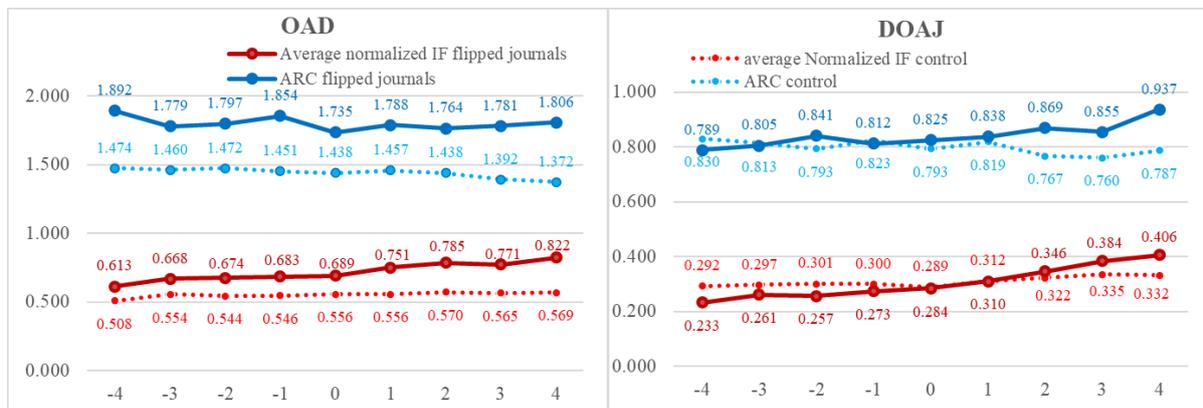

**Figure 4. Mean normalised IF (blue lines) and ARC (red lines) of OAD (left) and DOAJ (right) journals considered in this study. Solid lines represent the group of flipped journals, dotted lines the respective control group.**

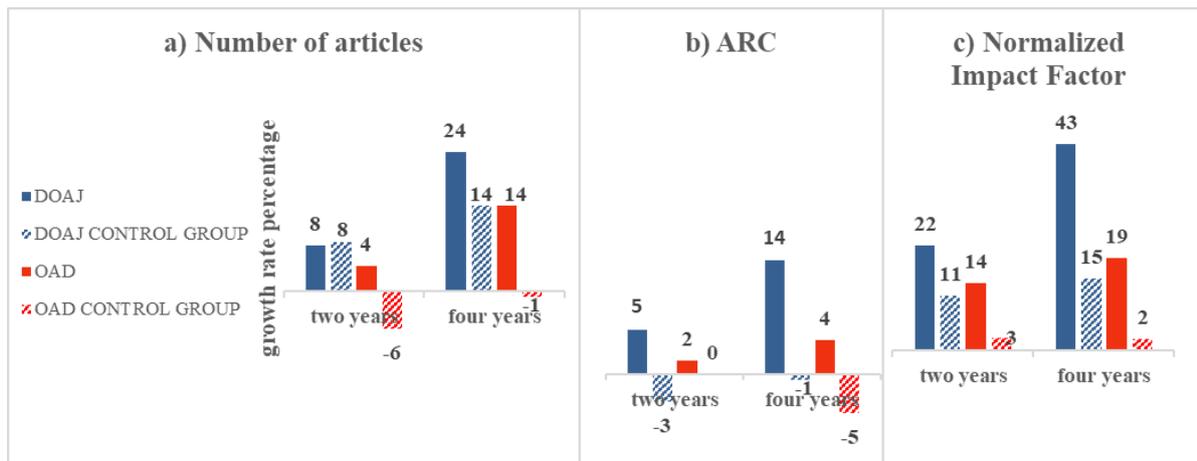

**Figure 5. Growth rates, calculated as the percentage growth between the flipping year and measurement year, for a) number of articles, b) ARC) and c) normalized impact factor. Growth rates were measured for both groups of flipped journals (DOAJ and OAD), as well as their respective control groups, at two and four years post-flipping.**

In the DOAJ group, we observe clearer temporal trends in ARC and normalized IF which may, at least in part, be attributed to the flipping of the journal. ARC increases from 0.83 to 0.93 between the flipping year and 4 years after the flipping year (an increase of 14%), whilst normalized IF increases from 0.29 to 0.41 (an increase of 43%). These values are both higher than those observed for the control group, which decreased ARC by 1% and increased normalized IF by 15%.

The effect of flipping a journal to OA on ARC and IF may be manifested differently across different scientific fields. To investigate these possible differences, we additionally grouped journals by field and compared changes in ARC and normalized IF between the time of flipping, and two years and four years post-flipping. Results for each field are shown in Table 2.

| #journals per field | | growth rate (%) two years after flip | | | | growth rate (%) four years after flip | | | |
|---|---|---|---|---|---|---|---|---|---|
| | | DOAJ | DOAJ Control group | OAD | OAD Control group | DOAJ | DOAJ Control group | OAD | OAD Control group |
| Health Sciences DOAJ: 117 OAD: 40 | #artic. | 10 | 9 | 7 | 1 | 21 | 13 | 15 | 9 |
| | Av. IF | 22 | 12 | 26 | 2 | 44 | 13 | 41 | 4 |
| | ARC | 6 | -6 | 1 | -3 | 14 | -7 | 2 | -10 |
| Social Sciences DOAJ: 49 OAD: 25 | #artic. | -7 | 2 | -3 | 2 | 43 | 4 | 7 | -2 |
| | Av. IF | 36 | 22 | 0 | 15 | 46 | 31 | 8 | 5 |
| | ARC | -4 | 7 | -1 | -14 | -5 | 4 | 16 | -11 |
| Life Sciences DOAJ: 71 OAD: 23 | #artic. | 4 | 18 | -3 | -7 | 16 | 36 | 13 | -6 |
| | Av. IF | 16 | 2 | 15 | -2 | 43 | 7 | 29 | 0 |
| | ARC | 8 | -7 | 2 | 0 | 20 | -4 | 3 | -4 |
| Physical Sciences DOAJ: 61 OAD: 24 | #artic. | 9 | 3 | -2 | -14 | 24 | 11 | 8 | -5 |
| | Av. IF | 29 | 12 | -5 | -2 | 48 | 15 | 0 | -2 |
| | ARC | 7 | -7 | 5 | 5 | 11 | 0 | 3 | 3 |

**Table 2. Two-year and four-year growth rates in numbers of published articles, normalized IF and ARC for CA and flipped journals, by scientific field. Numbers in the left column refer to the number of journals within each field, for DOAJ and OAD journals, respectively.**

In general, we observe a strong variability between the different bibliometric dimensions under study (i.e. number of articles published, normalised IF and ARC) and between each field, suggesting that the effect of flipping a journal differs strongly between different fields and included disciplines, respectively. For Health Sciences (117 journals in DOAJ and 40 journals in OAD), for example, growth rates of all dimensions were positive at two and four years after flipping for both sets of flipped journals, and higher than values observed in the control groups. Conversely, in the Social Sciences (49 journals in DOAJ, 25 journals in OAD), growth rates in the number of articles published are negative at two years after flipping, but become positive, and for DOAJ far greater than the growth rates of the control group, at four years after flipping, indicating that the effect of flipping, at least in terms of article volume, takes a longer time to diffuse in the Social Sciences.

## Conclusion

We have presented one of few studies on journals which flipped from a CA to an OA model and its effect on journal publication volumes, article- and journal-level citations metrics and how these compare to journals which still pursue the CA model. The literature reporting studies on flipped journals shows that journals' IFs usually increase after flipping (Busch, 2014a; Bautista-Puig et al., 2020). Our results agree with these previous findings, but show that whilst IF and ARC increase generally in the years following flipping, they vary greatly across scientific fields. Previous studies found that the effect of the OA model on received citations is field specific (Björk & Solomon, 2012; Li et al., 2018). One reason for the higher advantage by some disciplines is probably the lack of available OA journals at the same quality level for those disciplines. Of course, this effect is accelerated by the general relation between the quality of articles and journals and received citations which is not discipline-dependent. For example, Gargouri et al. (2010) found a greater OA advantage for articles published in journals with higher impact factors. Moreover, the amount of charged APCs may be a factor influencing the number of citations. Björk and Solomon (2012) showed that the average number of citations for OA journals with an APC model is higher than for those without an APC model. Zhang et al. (2020) found that 'Life Sciences' charge the highest APCs followed by 'Health Sciences' and 'Physical Sciences', while the 'Social Sciences' charge the lowest APCs.

We also observe that a higher number of articles are published after flipping, pointing to either a higher tendency amongst authors to submit to OA journals, which complements the research by Rowley et al. (2017), or higher acceptance rates by journals (e.g. because of lack of space-wise restrictions for online-only publications). Here again we saw different trends across fields for both groups of flipped journals. The willingness to submit to OA journals with APCs is related to the amount of available funding (e.g. from institutions, universities, or governments). Zhang et al. (2020) reported that authors from the 'Social Sciences' show a lower willingness-to-pay for APCs because of less financial support whereas authors from 'Health and Life Sciences' are able and willing to spend more on OA publishing because of more financial resources available.

Our study as well as related work have shown that flipping to an OA publishing model can positively affect the number of published articles as well as journal and article citation indicators. However, journals that flip to OA are confronted with a complex net of interrelated factors that determine success or failure of the flipping procedure. More in-depth studies are needed to control for the various factors affecting journal success.

## Limitations and Future work

This study has a number of limitations, which can be built upon and improved in future work. Most importantly, the study has a relatively small sample size, with only 234 journals considered from DOAJ, and 87 journals from OAD. It is therefore not clear how representative this sample is of the total number of journals that have flipped from CA to OA models – but it is almost certainly not a complete list of the entirety of flipped journals. Thus, more advanced methods for identifying journals that have flipped from CA to OA models (e.g. by utilising data from large-scale aggregators of OA information such as Unpaywall or CORE) may help to generate a more complete picture in future.

However, it should be noted that Unpaywall data might be a source of error also affecting our study (regarding the construction of the control groups). Akbaritabar and Stahlschmidt (2019) have studied Unpaywall and showed that 13% of publications that Unpaywall classified as OA was classified as CA in Crossref. Here more work is needed to better determine the OA/CA-status of articles.

Another limitation is the lack of data on submissions to flipped journals which we assume to better reflect the willingness of authors to publish in an OA journal. The results of our analyses are only based on the number of accepted articles which may have also increased due to changes in editorial policies, amongst others.

Although we used article and journal level citation indicators to increase the precision of comparisons of groups of flipped journals with CA journals we didn't exclude outliers at article level (e.g. the merit or quality of individual articles which might draw attention and higher impact among the community) and journal level (impact factor) which could affect the measures. Future work will be more sensible to such issues (e.g. by removing outliers from analyses).

A number of additional factors are important to consider, when using bibliometric indicators to understand the development of a flipped journal over time. For example, it is important to consider the exact business model that is used by a flipped journal – some journals may use an APC-driven model, whilst others may be supported by individual societies or library presses. These different models bring different economic challenges, as highlighted by Matthias et al. (2019) who found that a large percentage of journals that flipped to an OA model eventually flipped back to a CA model, in part for monetary reasons. These economic pressures may also cause downstream changes on editorial decisions, not least because APC-driven revenues are closely tied to journal acceptance rates. A related factor is that of the publisher itself – in this study we considered changes at the journal level, but did not consider how a change in the publisher may also accompany a change in business model. Different publishers bring differences in platform quality and visibility, and these may also have an effect, for example, on the willingness of authors to publish their work with a journal. Bautista-Puig et al. (2020) also investigated how countries of publishing and citing authors change before and after a flip, which is important for understanding exactly who is supporting new OA models and what effect that may have on bibliometric indicators and publishing behaviour. In addition, long-term changes in the support of institutions and funders, as well as increasing pressure to transition to OA models will mean that the findings presented here will evolve over time. Future work should therefore focus on trying to understand the complicated interactions between these different factors.

It is important that quantitative bibliometrics, such as the results presented here, also involve the views of stakeholders such as publishers, funders, libraries and societies. Therefore, future work should also be complemented with more qualitative information from interviews with these stakeholders, to reveal their attitudes towards journal flipping and OA, their expectations regarding journal quality and indicators as well as their motivation to change the publication model.

## Acknowledgement

This work is supported by BMBF project OASE, grant number 01PU17005A. We are thankful to Masoud Davari for his assistance with preparing and validating the data. We thank Najko Jahn and the anonymous reviewers for helpful comments on the manuscript.